\documentclass[letterpaper]{article} 
\usepackage{aaai24}  
\usepackage{times}  
\usepackage{helvet}  
\usepackage{courier}  
\usepackage[hyphens]{url}  
\usepackage{graphicx} 
\usepackage{natbib}  
\usepackage{caption} 
\usepackage{algorithm}
\usepackage{algorithmic}

\usepackage{newfloat}
\usepackage{listings}
\DeclareCaptionStyle{ruled}{labelfont=normalfont,labelsep=colon,strut=off} 
\floatstyle{ruled}
\newfloat{listing}{tb}{lst}{}
\floatname{listing}{Listing}
\usepackage{booktabs}
\usepackage{pifont}
\usepackage{cleveref}
\usepackage{amsfonts}
\newcommand{\xmark}{\ding{55}}%
\newcommand{\cmark}{\ding{51}}%
\usepackage{xcolor}

\newcounter{ff}
\def\google{%
      \ifnum\value{ff}=0%
        \footnote{Work done while at Google.}%
        \setcounter{ff}{\value{footnote}}%
      \else%
        \footnotemark[\value{ff}]%
      \fi%
    }

\title{V2Meow: Meowing to the Visual Beat via Video-to-Music Generation}
\author{
    Kun Su\equalcontrib\google\textsuperscript{\rm 3},
    Judith Yue Li\equalcontrib\textsuperscript{\rm 1},
    Qingqing Huang\google\textsuperscript{\rm 4},
    Dima Kuzmin\textsuperscript{\rm 1},
    Joonseok Lee\textsuperscript{\rm 1,5},\\
    Chris Donahue\textsuperscript{\rm 2,6},
    Fei Sha\textsuperscript{\rm 1},
    Aren Jansen\textsuperscript{\rm 1},
    Yu Wang\google\textsuperscript{\rm 7},
    Mauro Verzetti\textsuperscript{\rm 2},
    Timo Denk\textsuperscript{\rm 2}
}
\affiliations{
    \textsuperscript{\rm 1}Google Research, Mountain View, CA, USA \\
    \textsuperscript{\rm 2}Google DeepMind, Mountain View, CA, USA \\
    \textsuperscript{\rm 3}University of Washington, Seattle, WA, USA \\
    \textsuperscript{\rm 4}ByteDance, San Jose, CA, USA \\
    \textsuperscript{\rm 5}Seoul National University, Seoul, South Korea \\
    \textsuperscript{\rm 6}Carnegie Mellon University, Pittsburgh, PA, USA \\
    \textsuperscript{\rm 7}Music and Audio Research Laboratory, New York University, NY, USA \\
    \{judithyueli, dimakuzmin, joonseok, chrisdonahue, fsha, arenjansen, verzetti, timodenk\}@google.com, suk4@uw.edu

}

\begin{document}

\maketitle

\begin{abstract}
Video-to-music generation demands both a temporally localized high-quality listening experience and globally aligned video-acoustic signatures. While recent music generation models excel at the former through advanced audio codecs, the exploration of video-acoustic signatures has been confined to specific visual scenarios. In contrast, our research confronts the challenge of learning globally aligned signatures between video and music directly from paired music and videos, without explicitly modeling domain-specific rhythmic or semantic relationships. We propose V2Meow, a video-to-music generation system capable of producing high-quality music audio for a diverse range of video input types using a multi-stage autoregressive model. Trained on 5k hours of music audio clips paired with video frames mined from in-the-wild music videos, V2Meow is competitive with previous domain-specific models when evaluated in a zero-shot manner. It synthesizes high-fidelity music audio waveforms solely by conditioning on pre-trained general-purpose visual features extracted from video frames, with optional style control via text prompts. Through both qualitative and quantitative evaluations, we demonstrate that our model outperforms various existing music generation systems in terms of visual-audio correspondence and audio quality. Music samples are available at tinyurl.com/v2meow.
\end{abstract}

\section{Introduction}
\label{sec:intro}
\begin{figure*}
    \centering
    \includegraphics[width=0.85\linewidth]{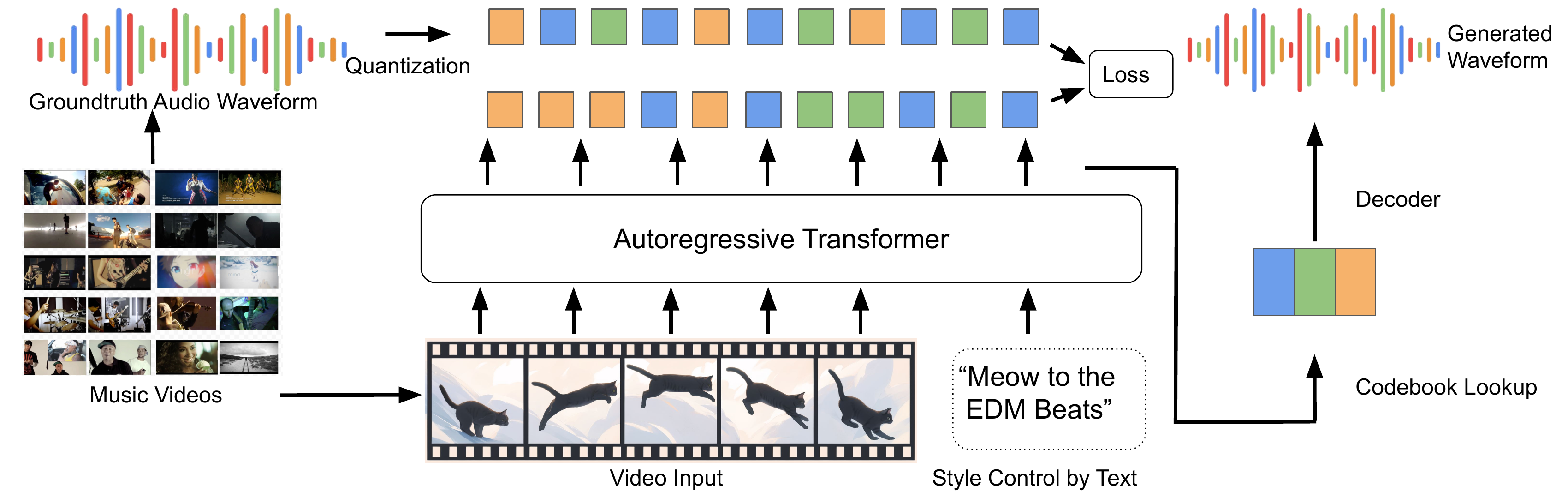}
    \caption{The video-to-music generation model V2Meow synthesizes high-fidelity music conditioned on video input and optionally text describing high-level style.}
    \label{fig:teaser}
\end{figure*}

Recent advancements in high-resolution neural audio codecs~\cite{zeghidour2021soundstream, Defossez2022HighFN, kumar2023highfidelity} have introduced novel possibilities for directly generating high-quality music waveforms comparable to human-made music~\cite{borsos2022audiolm, Agostinelli2023MusicLMGM}. Notably, AudioLM~\cite{borsos2022audiolm} employs a multi-stage autoregressive modeling approach for audio generation. It utilizes a masked language model pre-trained on tokenized audio encodings~\cite{chung2021w2v} to capture long-term structures and the discrete neural audio codec~\cite{zeghidour2021soundstream} for high-quality synthesis. Music language models like MusicLM~\cite{Agostinelli2023MusicLMGM} have further demonstrated that autoregressive models can generate music conditioned on text prompts~\cite{Agostinelli2023MusicLMGM,copet2023simple} or input vocals~\cite{donahue2023singsong}. However, the challenge persists in video-conditioned music generation, where a gap exists between the coarser global video signature and the aforementioned high-resolution audio representations or audio codecs for waveform generation.

Most existing work on video-to-music generation focuses on modeling the audiovisual correspondence between domain specific video representation and symbolic representation of music. For example, dance-to-music literature~\cite{Zhu2022QuantizedGF, Zhu2022DiscreteCD, Yu2023LongTermRV} relies on modeling music rhythms, style from annotated human motion, other works focuses on generating natural sound that is faithful to the physical motion in the silent video~\cite{su2020audeo, FoleyMusic2020,owens2016visually,zhou2018visual, chen2020generating,su2023physics}, for example, reconstructing instrumental music from silent instrument performance videos. As a result, the input video types are restricted to certain visual scenarios, and cannot be generalized to arbitrary video input types, e.g., a cat video or slideshows of images. In contrast, we propose to learn a general audiovisual correspondence directly from paired video frames and music waveform data. Specifically, we bridge the gap between the coarser video representation and the high-resolution audio representation through finding the video-audio aligned low-resolution representation space. This approach allows us to generalize to a wide range of video input types, and leverage the vast amount of parallel music and video data available on the internet for scaling.

We introduce \emph{V2Meow}, a high-fidelity music audio waveform generator conditioned on diverse video inputs. Drawing inspiration from MusicLM~\cite{Agostinelli2023MusicLMGM}, we employ a multi-stage autoregressive language modeling approach. V2Meow takes video frames as input with optional style control through text prompts, treating video and text as a unified input stream fed into the Transformer with feature-specific adaptors. By not explicitly modeling domain-specific audiovisual correspondence, V2Meow exhibits zero-shot transfer capabilities, demonstrated in evaluations on AIST++ dance videos~\cite{li2021ai}. Quantitative and qualitative assessments of music-video correspondence, along with an extensive ablation study, elucidate factors influencing generation quality. Numerical and human study results affirm that, compared to MIDI-based baselines, V2Meow aligns better with human music preferences. Additionally, including video input enhances V2Meow's visual relevance learning compared to text-only input.

\section{Related Work}
\label{sec:related}
\noindent\textbf{Video to Audio.}
Over the past few years, there have been advancements in deep-learning approaches to produce realistic sounds from silent videos.
\citet{owens2016visually} initially predicted impact sound features based on image features and then retrieved the closest sound sample from the dataset instead of directly generating the sound.
\citet{chen2017deep} investigated the utilization of conditional GAN for generating sound from images, although the experiments were constrained to music performances collected in a laboratory setting.
Visual2Sound~\cite{zhou2018visual} suggested generating audio waveform from videos captured in-the-wild, but the dataset is limited to only ten types of sounds. Subsequently, novel loss functions~\cite{chen2018visually, chen2020generating} are introduced to enhance the semantic alignment of generated audio from videos. The effectiveness of models for generating audio from videos is primarily constrained by the typically weak correspondence between the video and audio, as well as the limited scale of training data. Generating high-quality audio from a silent video poses a challenge without a robust audio generative model and adequate audio representations.

\noindent\textbf{Video to Music.}
Apart from natural sounds, several studies have delved into the generation of music from videos. Initial efforts focused on generating symbolic music (MIDI) from videos depicting a musician playing the piano \cite{koepke2020sight,su2020audeo} or other instruments~\cite{FoleyMusic2020, su2020multi}. Later, RhythmicNet~\cite{su2021does} showcased the potential to generate music soundtracks synchronized with arbitrary human body movements. Following studies have expanded the generation of symbolic music representations to encompass waveform generation~\cite{Zhu2022DiscreteCD,Zhu2022QuantizedGF,Yu2023LongTermRV}. Nevertheless, the music generation systems proposed still depend on intricate motion extractors to explicitly capture visual rhythm from domain-specific data, such as dance videos. Apart from relying on visual cues from human motion, a music Transformer has been suggested for the generation of video background music~\cite{di2021video}, albeit through MIDI generation. All these approaches face challenges due to their domain-specific modeling assumptions, such as focusing on music from particular instruments or relying heavily on visual cues like human body motion. Consequently, the generated samples are constrained to specific instrument types and visual scenarios. In contrast to the aforementioned studies, our approach utilizes music videos captured in real-world settings to establish a general mapping from visual input to audio waveforms.

\noindent\textbf{Music Generation.}
A robust music representation such as MIDI has been widely employed in the modeling of music. Early studies transformed MIDI into piano-roll representation by employing GANs~\cite{dong2018musegan} or variational autoencoders~\cite{roberts2018hierarchical, groove2019} to generate novel music.
Subsequently, there were proposals for event-based representations aimed at more efficient representation of MIDI~\cite{oore2020time,huang2018music,hawthorne2018enabling, huang2020pop}. Control signals are additionally integrated into the music generative models based on MIDI~\cite{engel2017neural, choi2019encoding, lattner2019high}.

In terms of modeling music directly from raw audio without the need for transcripts or symbolic music representations, WaveNet~\cite{oord2016wavenet} introduced autoregressive modeling to synthesize music audio with satisfactory quality. Jukebox~\cite{dhariwal2020jukebox} adopted a hierarchical approach to generate tokens at different temporal resolutions, which were subsequently combined to reconstruct music. Recent work on high quality audio representation~\cite{zeghidour2021soundstream,Defossez2022HighFN,kumar2023highfidelity} directly apply residual vector quantization on the raw waveform. Later, several recent works adopted such representation for text-to-audio generation using transformer-based autoregressive models, e.g., AudioLM~\cite{borsos2022audiolm}, Mubert~\cite{mubert2022}, AudioGen~\cite{kreuk2022audiogen} and MusicLM~\cite{Agostinelli2023MusicLMGM} or non-autoregressive models~\cite{copet2023simple,garcia2023vampnet}. Alternatively, Riffusion~\cite{Forsgren_Martiros_2022} and other recent work~\cite{Huang2023Noise2MusicTM,liu2023audioldm,huang2023makeanaudio,Schneider2023MosaiTG} adopted a diffusion based approach.

\section{The Proposed Method: V2Meow}
\label{sec:method}

\begin{figure*}[!t]
    \centering
    \includegraphics[width=1.0\textwidth]{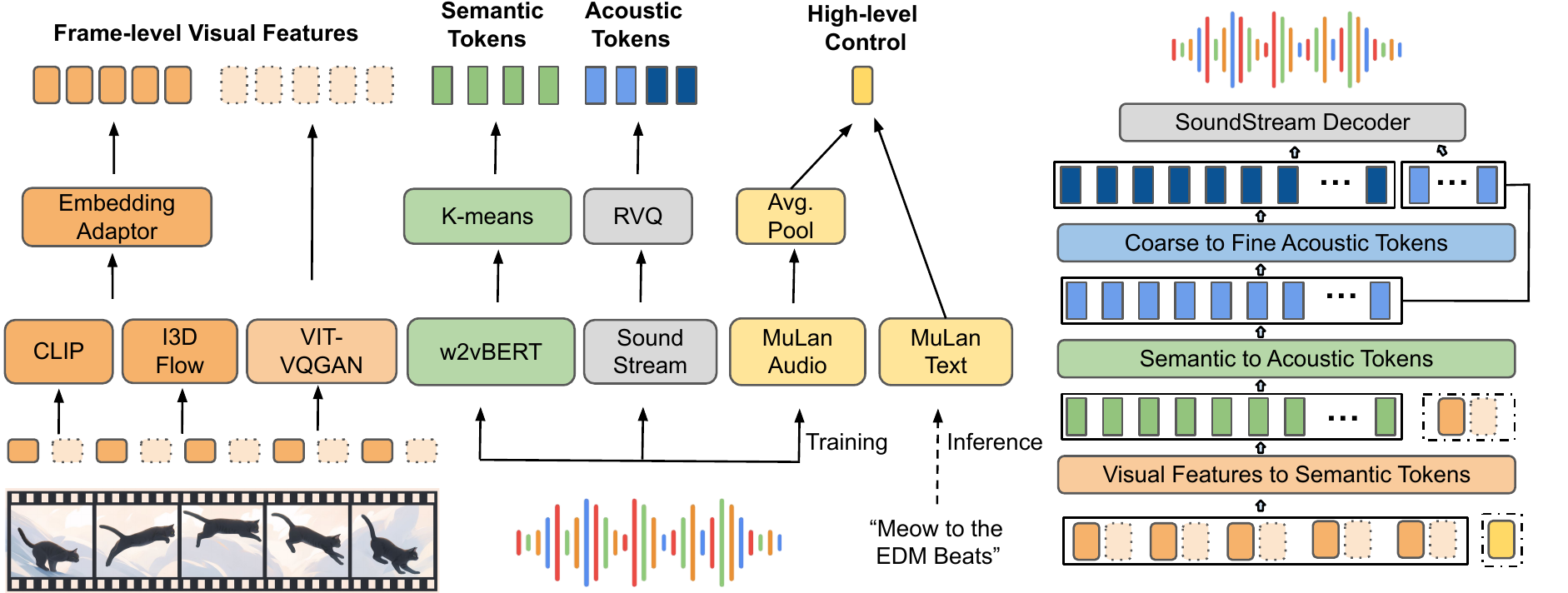}
    \caption{V2Meow Architecture Overview: (left) Feature extraction pipeline for video, audio and text representations. (right) Overview of multi-stage video to music modeling. }
    \label{fig:v2meow_overview}
\end{figure*}

In this section, we describe the feature representations and modeling pipeline of our proposed method, V2Meow, in detail.

\subsection{Feature Representations}
\label{sec:method:feature}

For audio waveforms, we follow MusicLM~\cite{Agostinelli2023MusicLMGM}, using semantic and
acoustic tokens extracted from two different pre-trained self-supervised models.
For visual inputs, we explore various types of visual features to find the most informative representation suitable for music generation task.
Finally, we demonstrate how we represent the control signal without paired music-video-text examples.

\noindent\textbf{Semantic Music Tokens.}
We extract semantic tokens using a pre-trained w2v-BERT model~\cite{chung2021w2v}.
w2v-BERT is a self-supervised model using masked language modeling (MLM) with contrastive loss, coarsely learning to represent audio, capturing both local dependencies such as local melody in music and global long-term structure such as harmony and rhythm.
To obtain the semantic tokens, we extract embeddings from an intermediate layer of w2v-BERT.
We then apply $k$-means algorithm with $K_s$ clusters on these embeddings and use the centroid indices as semantic tokens.
For each audio waveform, we obtain semantic tokens $\{S_t : t = 1, ..., T_s\}$, where $T_s$ is the total number of tokens.
While the coarse resolution of the semantic tokens enables us to model long-term dependencies, the audio reconstruction solely from these semantic tokens usually leads to poor quality.

\noindent\textbf{Acoustic Music Tokens.}
To generate high-quality music audio, we additionally rely on acoustic tokens extracted from a pre-trained SoundStream~\cite{zeghidour2021soundstream} model.
SoundStream is a universal neural audio codec that compresses arbitrary audio at low bit rates and reconstructs the audio back in a high quality.
Specifically, a convolutional encoder embeds the input waveform, followed by residual vector quantization (RVQ) to discretize them.
RVQ is a hierarchical quantization scheme composing a series of $N$ vector quantizers, where the target signal is reconstructed as the sum of quantizer outputs.
Each quantizer with a vocabulary size of $K_a$ learns to quantize the embedding simultaneously during training. Thanks to the residual quantization, the acoustic tokens have a hierarchical structure such that tokens from the coarse quantizers recover acoustic properties like music recording conditions, while leaving only the fine acoustic details to the fine quantizer tokens.
Since coarser levels are more important for high-fidelity reconstruction as illustrated in AudioLM, we first construct a mapping from music semantic tokens to coarse acoustic tokens and then learn a mapping from coarse acoustic tokens to fine-grained acoustic tokens in the later stage.

\subsection{Model Conditioning}
\noindent\textbf{Video Frames Conditioning.}
Given a video as a sequence of $T$ frames, $\left\{\mathbf{v}_t \in \mathbb{R}^{H \times W \times 3} : t = 1, ..., T\right\}$, we aim to extract useful visual features from existing pre-trained visual model.
We explore various visual representations for this, among pure visual models, multimodal models, and quantized models,
since it is unclear which kind of visual representation could provide sufficient information for music generation. In particular, we explored a combination of the following visual features as.
\begin{enumerate}
  \setlength{\itemsep}{0pt}
  \setlength{\parskip}{0pt}
  \item \textit{Purely visual representations}: Video understanding models learn underlying patterns from the pixel distributions observed in a collection of images or videos, using CNNs~\cite{tran2015learning,carreira2017quo,feichtenhofer2020x3d} or Transformers~\cite{arnab2021vivit,bertasius2021space}, without access to additional modalities.
  As it is common that the visual changes in a video have correspondences to musical rhythm, we adopt the Inflated 3D (I3D), which explicitly considers the optical flow, which is known to be useful for analyzing motions. In our experiments, we denote the visual flow embeddings as $\left\{\mathbf{f}_t \in \mathbb{R}^{D_f} : t = 1, \dots, T\right\}$, extracted from an I3D model pretrained on Kinetics~\cite{carreira2017quo}, where $D_f$ indicates the dimensionality of the I3D features.
   
  \item \textit{Multimodal embeddings}:
  The second type is an embedding learned from multimodal correspondence, in addition to the visual modality. Contrastive Language Image pre-training (CLIP)~\cite{radford2021learning} is a popular image-text model, widely used in a variety of downstream tasks including generative applications. We expect its generality and robustness would be potentially useful to incorporate semantics of the video. We denote the CLIP embeddings as $\{\mathbf{c}_t \in \mathbb{R}^{D_c} : t = 1, \dots, T\}$, where $D_c$ is the dimensionality of the CLIP embedding.
  
  \item \textit{Visual tokens}: Since the semantic and acoustic music representations in our pipeline are both discrete tokens, we explore using a similar type of discrete tokens for visual inputs. To obtain discrete tokens for a video frame, we adopt ViT-VQGAN~\cite{yu2021vector}, the state-of-the-art self-supervised Vision Transformer~(ViT) model that performs image quantization on each image to obtain a set of discretized latent codes and uses a Transformer to predict these image tokens autoregressively for image reconstruction. Given a video frame $\mathbf{v}_t \in \mathbb{R}^{H \times W \times 3}$, where $H, W$ indicate the image height and width, respectively, the VIT-VQGAN encodes the image into $H/D_q \times W/D_q$ discretized latent codes, where $D_q$ is the size of non-overlapping image patches mapped to one token. A video with $T$ frames is represented as a set of tokens $\left\{Q_t \in \mathbb{Z}_+^{H/D_q \times W/D_q} : t = 1, \dots, T\right\}$.
\end{enumerate}

\noindent\textbf{Text Conditioning.}
Since the music for a video could be highly dependent on personal preference, we allow users to optionally provide a music-related text description, in addition to the visual input, to control the generated music at a high-level. However, it is challenging to collect music-video-text pairs in the wild.
To overcome this issue, we leverage a music-text joint embedding model, MuLan~\cite{huang2022mulan}, which is trained on paired music-text data using a contrastive loss.
For each video in the training set, V2Meow first extracts MuLan embeddings of all audio segments $\left\{\mathbf{m}_j \in \mathbb{R}^{D_m} : j = 1, \dots, J\right\}$, where each segment is ten second long and $J$ indicates the total number of audio segments in the video, $D_m=128$ is the dimension of MuLan audio embedding.
We then average the embeddings into a single video-level.
It is worth noting that we extract a fixed-length segment at a random starting point from a music video at each training iteration.
Although it would be ideal to perform inference on MuLan with the selected segment only, we avoid doing this aiming for an efficient experiment. Instead, we use a video-level embedding for the entire video and empirically verify that this is sufficient, probably because our goal is to have a high-level control on the style of generated music, instead of fine-grained control. At inference time, we may use a music-related text description to obtain a MuLan embedding and condition our V2Meow model on it.

\subsection{Modeling Pipeline}
\label{sec:method:pipeline}
We adapt the AudioLM pipeline to train the visual conditioned music generation model. There are three main stages of sequence-to-sequence modeling tasks.

\noindent\textbf{Stage 1. Visual Features to Music Semantic Tokens.}
In the first stage, V2Meow learns a mapping from visual inputs to the music semantic tokens.
Specifically, we use an encoder-decoder Transformer~\cite{vaswani2017attention} where the encoder takes the visual features, and the decoder predicts the music semantic tokens autoregressively.
It turns out that this stage is the most critical part of generating music which reflects the video well.
On the one hand, it builds up the connection between visual and audio modalities, and models the semantic transformation from visual information to audio.
On the other hand, this stage does not output high-quality fine-grained audio, allowing the model to focus on associating the two modalities with each other.

\noindent\textbf{Stage 2. Music Semantic Tokens to Coarse Acoustic Tokens.}
In the second stage, we aim to convert the music semantic tokens to acoustic tokens for high-quality synthesis.
We follow the AudioLM pipeline to split this stage into coarse and fine acoustic modeling.
In the coarse acoustic modeling, we explore two different training strategies:
1)~We follow AudioLM to train a decoder-only Transformer to map music semantic tokens to coarse acoustic tokens.
Since such a training strategy does not require visual information, we could use the large-scale music data in the wild to pretrain a robust model.
2)~We also explore to see whether adding visual conditioning at this stage improves the performance.
Specifically, we train an encoder-decoder Transformer model where the encoder takes the visual features and music semantic tokens and the decoder generates the coarse acoustic tokens.
While ideally the second approach should work better, the final results are not necessarily better since the amount of music-video pairs available is still incomparable to the amount of music-only data.

\noindent\textbf{Stage 3. Coarse to Fine Acoustic Tokens and Audio Decoding.}
Once we have the coarse acoustic tokens, we follow AudioLM~\cite{borsos2022audiolm} and perform coarse to fine acoustic tokens modeling.
This stage maps the tokens in the first $N_c$ levels of SoundStream RVQ to the tokens of the remaining $N_f$ levels.
Finally, all levels of tokens are passed to SoundStream Decoder to reconstruct the audio.

\noindent\textbf{Adding Control.}
To incorporate the control signal into V2Meow during training, we simply feed the MuLan audio embedding as an additional input with a sequence length be one to the Transformer encoder along with the visual features in the first stage. Both Mulan audio embedding and visual features are projected to the same feature dimension.
At inference, we instead use the MuLan text embedding with the visual features to generate the semantic tokens.

\section{Experiments}
\label{sec:exp}

\subsection{Experimental Settings}
\label{sec:exp:setting}

\noindent\textbf{Training Datasets.}
Following~\cite{suris2022s}, we filtered a public available video dataset~\cite{abu2016youtube} to 110k videos with the label Music Videos and refer to it as MV100K. The training and validation datasets were split into an 80:20 ratio. We trained the Stage 1 model and Stage 2 model on 5k hours of music videos. A version of Stage 2 model is trained on audio-only data for ablation study.
For computing semantic and acoustic audio tokens, we adopt the SoundStream tokenizer and w2v-BERT tokenizer, both of which are pre-trained on 46k hours of music only audio data sampled at 16kHz sampling rate.

\noindent\textbf{Evaluation Datasets.} 
We evaluate our methods on three different datasets. 
For the task of video conditional music generation, we use the test partition of the MV100K. We select 13~genres of music videos to comprise a genre balanced subset with a total number of 4076~videos. 
For the task of video and text conditional music generation, we use the latest MusicCaps dataset~\cite{Agostinelli2023MusicLMGM} which is a subset of AudioSet~\cite{gemmeke2017audio}. The MusicCaps has about 5.5k~human annotated text captions, music, and video pairs. With the text caption, we can verify whether the generated music could be controllable and whether its performance is comparable with text-to-music generation models like MUBERT~\cite{mubert2022} and Riffusion~\cite{Forsgren_Martiros_2022}. For both tasks, we generate ten~second audio for each video clip. 
For dance-to-music generation task, we evaluate temporal alignment on 20 dance videos in the test split of AIST++~\citep{li2021ai}. The evaluation is in zero-shot fashion without any fine-tuning on the AIST++ train split, and only video frames are used for modeling while no motion data is involved. The reported metrics represent averages over 20 10-second audio segments and 86 2-second audio segments, with 5 inference examples per segment.

\noindent\textbf{Implementation Details.}
For all visual features, we use a frame rate at $1$~fps, following the standard on MV100K~\cite{abu2016youtube}.
We use the released ViT-L/14 model\footnote{huggingface.co/sentence-transformers/clip-ViT-L-14} to extract the CLIP embeddings, whose dimensionality is $768$.
For computing the I3D~Flow embeddings, we use a model pre-trained on the Kinetics dataset, whose dimensionality is $1024$.
We use a pre-trained VIT-VQGAN encoder to obtain $1024$ tokens for each image and the vocabulary size is $8192$.
For the visual feature to music semantic tokens modeling, we use encoder-decoder Transformer with 12~layers, 16~attention heads, an embedding dimension of $1024$, feed-forward layers of dimensionality $4096$, and relative positional embeddings.
We use 10-second random crops of the music video for visual to music semantic tokens modeling and semantic tokens to coarse acoustic tokens modeling.
The coarse to fine acoustic tokens modeling is trained on 3-second crops. During inference, we use temperature sampling for all stages, with temperatures $\{1.0, 0.95, 0.4\}$ for modeling stages 1, 2, and 3, respectively.

\subsection{Evaluation Metrics}
\label{sec:exp:numerical}

\begin{table*}[htb!]
  \begin{center}
  \small
  \begin{tabular}{lcc|cccc|p{1.6cm}p{1.6cm}}
    \toprule
 &&&\multicolumn{4}{c}{Semantic / Semantic + Acoustic Modeling} &\multicolumn{2}{c}{Semantic + Acoustic Modeling}\\
Method    &Visual   &Text & FAD TRILL~\textcolor{black}{$\downarrow$} & FAD VGG~\textcolor{black}{$\downarrow$}  & KL Div.~\textcolor{black}{$\downarrow$} & MCC~\textcolor{black}{$\uparrow$} & Visual Relevance~\textcolor{black}{$\uparrow$} & Music Preference~\textcolor{black}{$\uparrow$} \\ \midrule
\multicolumn{3}{l|}{\underline{\textit{Eval Dataset}}: MV100K} & & & & \\
CMT &\cmark & \xmark & N/A & N/A & N/A & N/A & 20.6\% & 30.0\%\\
Random Shuffle & \xmark    &\xmark &- &- &0.67 &0.268 & N/A & N/A\\
V2Meow-CLIP     & \cmark    &\xmark &0.236/0.158 &6.094/2.779 &0.63/0.54 &0.312/0.372 & 78.2\% & 67.6\% \\
V2Meow-I3D & \cmark &\xmark  &0.236/\textbf{0.151} &6.278/2.328 &0.77/0.65  &0.279/0.296 & 74.3\% & 65.8\% \\
V2Meow-VIT & \cmark &\xmark &0.240/0.174 &6.097/1.988 &0.73/0.62 &0.276/0.294 & 81.4\% & 71.5\%\\
V2Meow-VIT+I3D	 &\cmark  & \xmark  &0.236/0.178 & 5.801/\textbf{1.945} &0.68/0.57 &0.298/0.327 & 83.8\% & 76.8\%\\
V2Meow-CLIP+I3D	 &\cmark  & \xmark & 0.235/0.165    &6.126/2.003   & 0.64/\textbf{0.49} & 0.343/\textbf{0.419} & 79.2\% & 68.2\% \\ \hline
\multicolumn{3}{l|}{\underline{\textit{Eval Dataset}}: MusicCaps} & & & &\\
CMT &\cmark & \xmark & N/A & N/A & N/A & N/A & 19.7\% & 20.7\%\\
Riffusion &\xmark & \cmark & 0.760 & 13.4 & 1.19 & 0.34 & 38.6\% & 41.2\%\\
MUBERT  &\xmark & \cmark  &0.450 &9.6 &1.58 & 0.32 & 43.3\% & 49.3\% \\
V2Meow-CLIP     & \cmark    &\cmark &0.379/\textbf{0.328} &5.198/4.628 &1.31/1.19 &0.364/0.377 & 63.6\% & 58.5\% \\
V2Meow-I3D & \cmark &\cmark  &0.389/0.331 &5.190/\textbf{4.623} & 1.26/1.22 &0.377/0.371 & 68.8\% & 68.0\% \\
V2Meow-VIT & \cmark &\cmark &0.377/0.366 &4.970/5.039 & 1.34/1.23 &0.380/0.392 & 66.9\% & 60.3\% \\
V2Meow-VIT+I3D	 &\cmark  & \cmark  &0.381/0.359 & 5.094/4.819 &1.34/1.21 &0.379/0.389 & 71.8\% & 67.1\%\\
V2Meow-CLIP+I3D	 &\cmark  & \cmark & 0.391/0.349    &5.385/4.948   & 1.27/\textbf{1.19} & 0.369/\textbf{0.394} & 67.4\% & 65.8\%\\ \bottomrule
 \end{tabular}
  \end{center}
    \caption{Quantitative evaluations on MV100K and MusicCaps for different models. For FAD and KL Divergence, lower is better. For MCC, higher is better. Bold font indicates the best value. The five V2Meow variants are named based on the video features used as input. Semantic Modeling indicates video conditioning is used only for semantic modeling, while Semantic + Acoustic Modeling indicates video conditioning is used for both semantic and acoustic modeling.}
  \label{tab:quantitative results}
  \end{table*}

\begin{table}[tp]
\centering
\small
{%
    \renewcommand{\tabcolsep}{5pt}
\begin{tabular}{l|ccc}
\toprule
Model  (Length) & Beat Coverage &  Beat Hit & F1\\
\midrule
GT & 100                                 & 100  & 100\\
V2Meow CLIP+I3D (10s)   & 100.0 (0.00)                                 & 84.4 (25.1)               & 91.5              \\
V2Meow CLIP+I3D (6s) & 99.3 (8.64) & 84.7 (25.7) & 91.4 \\
V2Meow CLIP+I3D (2s)   & 90.0 (30.0)                              & 84.8 (32.1)                &     87.3        \\
CDCD Step-Intra (6s)  & 87.9                                 & 83.2                 &     85.5       \\
D2M-GAN (2s)  & 88.2                                 & 84.7                         &    86.4\\
CMT (2s) & 85.5                                 & 83.5                      &  84.5   \\ \bottomrule
\end{tabular}%
}
\caption{Zero-shot evaluation results on AIST++. For CMT and V2Meow only video frames are used as input, while CDCD Step-Intra and D2M-GAN requires additional motion annotation as inputs. For each video input we randomly generate $10$ music samples and report the average score and standard deviation.}
\label{tab:aist_eval}
\end{table}

\noindent\textbf{Objective Metrics.}
We follow~\cite{Agostinelli2023MusicLMGM} to use different quantitative metrics to automatically assess the fidelity, the semantic relevance of the generated samples and~\cite{Zhu2022QuantizedGF,Zhu2022DiscreteCD} to evaluate rhythmic alignment.
\begin{itemize}
    \item \textbf{Audio Quality.} We use Fréchet Audio Distance (FAD) based on two audio embedding models to measure different aspects of the audio quality, both of which are publicly available (1) TRILL\footnote{tfhub.dev/google/nonsemantic-speech-benchmark/trill/3}~\cite{shor2020towards}, which is trained on speech data, and (2) VGGish\footnote{tfhub.dev/google/vggish/1}~\cite{hershey2017cnn}, which is trained on the public audio event dataset~\cite{abu2016youtube,lee20182nd}.
    \item \textbf{Semantic Relevance.} KL Divergence (KLD)~\cite{yang2022diffsound, kreuk2022audiogen} and MuLan Cycle Consistency (MCC)~\cite{Agostinelli2023MusicLMGM} is used to determine whether generated music is semantically relevant to the reference audio or text. We run a LEAF classifier~\cite{zeghidour2021leaf} for multi-label classification on AudioSet, and use KLD over the predicted class probabilities between the original audio and the generated audio to evaluate if they share similar concept. For the video to music task, we use MuLan audio embedding of ground truth audio as reference to compute MCC as the average cosine similarity between the MuLan audio embedding of the generated music audio and reference. For the video and text to music task, we use the MuLan text embedding of the text description as reference instead to check text adherence. 
    \item \textbf{Rhythmic Alignment.}~\cite{Zhu2022QuantizedGF,Zhu2022DiscreteCD} introduces Beats Coverage Scores (BCS) and Beats Hit Scores (BHS) to count the aligned rhythm points of synthesized music and ground-truth music. BCS refers to the fraction of generated musical beats by the ground truth musical beats, while BHS refers to the ratio of aligned beats to the ground truth beats. Here we adopt the adjusted BCS and BHS introduced in~\citet{Yu2023LongTermRV} and compute F1 score in addition.
\end{itemize}

\noindent\textbf{Subjective Metrics.}
Whether the video and background music match is subjective. The generated music can be a reasonable match to the video, even if it is not similar to the ground truth music that accompanies the original video. Thus we conduct a human study to measure visual relevance and music preferences. Specifically, we sampled 89 distinct video examples from the MV100K test set and 76 distinct video examples from the MusicCaps test set. We surveyed around $200$ participants individually, and each participant was asked to evaluate a pair of videos with the same video but different background music. Each video pair is rated by $3$ person. A total number of $3500$ ratings are collected in the end.
\begin{itemize}
    \item \textbf{Visual Relevance}. We asked human raters to conduct a side-by-side comparison of the music generated from the baseline models (CMT, Riffusion, or MUBERT) and the five different V2Meow model variants by answering the question: "Which music do you think goes best with the video?". They are asked to ignore the sound quality and only focus on how well the music matches the video. 
    \item \textbf{Music Preference}. We asked human raters to choose which music they prefer to hear and ignore the video content. This task aim to study whether the generated music is aligned with human perceptual preference. Here we ask the listener to ignore sound quality and tell us which music they like.
\end{itemize}

\section{Results}
We initiate our evaluation of V2Meow's video-to-music generation capabilities by comparing it to the state-of-the-art model CMT~\cite{di2021video}, which relies on video-driven symbolic music representations. This evaluation is conducted on the MV100K dataset. Subsequently, we extend our comparison to two text-to-music systems, Mubert~\cite{mubert2022} and Riffusion~\cite{Forsgren_Martiros_2022}, using the MusicCaps dataset augmented with videos. The objective here is to assess the impact of incorporating video frames as conditioning signals in additional to text. For the task of dance-to-music generation, we further compare V2Meow with baseline models D2M-GAN~\citep{Zhu2022QuantizedGF}, CDCD Step-Intra~\citep{Zhu2022DiscreteCD}, and CMT~\citep{di2021video} on the AIST++ test split, aiming to evaluate V2Meow's understanding of complex dance motion. Detailed results are presented in Table~\ref{tab:quantitative results} and Table~\ref{tab:aist_eval}. The assessment concludes with an ablation study that dissects the significance of each stage in the modeling pipeline.

\subsection{Video Conditional Music Generation}
In the MV100K dataset, introducing video conditioning during the acoustic modeling stage notably enhances both audio quality-related metrics and semantic relevance, as illustrated in the second column of Table~\ref{tab:quantitative results}. Notably, when conditioned on specific visual embeddings, we find that clip embedding attains the highest MCC score, whereas I3D flow embedding exhibits superior performance in FAD metrics. This implies that different visual features capture distinct aspects within the video-music aligned subspace. The combination of Clip and I3D Flow embeddings achieves the highest MCC score across all models, with a corresponding enhancement in FAD VGGish compared to models with either Clip or I3D Flow embedding alone. While VIT-VQGAN tokens do not surpass others in individual metrics, the amalgamation of VIT-VQGAN tokens and I3D Flow embedding demonstrates improved performance compared to a single visual input. In terms of visual relevance and music preference, V2Meow significantly outperforms CMT by a substantial margin, as indicated in the third column of Table~\ref{tab:quantitative results}.
\subsection{Video and Text Conditional Music Generation}
In the MusicCaps evaluation, our approach, enhanced by the inclusion of video frames as an additional control, demonstrates a 20-30\% improvement in visual relevance compared to Riffusion and MUBERT that only condition on text. It's noteworthy that our approach also achieves lower FAD and higher MCC scores. Here the MCC is the similarity between generated music audio and text. This indicates that augmenting the conditioning with video frames not only enhances visual relevance but also contributes to improved audio quality and text adherence, despite utilizing a relatively small dataset of 5,000 hours of music videos. The combination of Clip and I3D Flow embeddings maintains the highest KLD and MCC scores, while the combination of VIT-VQGAN tokens and I3D Flow embeddings achieves the best visual relevance. Across all variations, V2Meow consistently outperforms the baseline in terms of audio quality, text adherence, visual relevance and music preference.
\subsection{Dance to Music Generation}
The zero-shot evaluation on dance videos in the AIST++ test split~\cite{li2021ai} reveals that V2Meow can attain performances comparable to those of specialized dance-to-music generation baselines~\cite{Zhu2022QuantizedGF, Zhu2022DiscreteCD}, measured by beat coverage and beat hit. This evaluation is conducted in a zero-shot fashion, without any fine-tuning on the AIST++ training split. Only video frames are used for modeling, and no human-annotated motion data is involved. The results suggest that our proposed framework can adeptly handle videos with significant rhythm changes, even when dance motion occurs below our 1fps sampling rate. It's important to note that D2M-GAN and CDCD Step-Intra utilize the AIST++ training split for fine-tuning, requiring a much higher sampling rate and additional motion annotation as input. In contrast, our model and CMT exclusively take video frames as input.

\subsection{Ablation Study} %
\begin{figure}[!t]
    \centering
    \includegraphics[width=\linewidth]{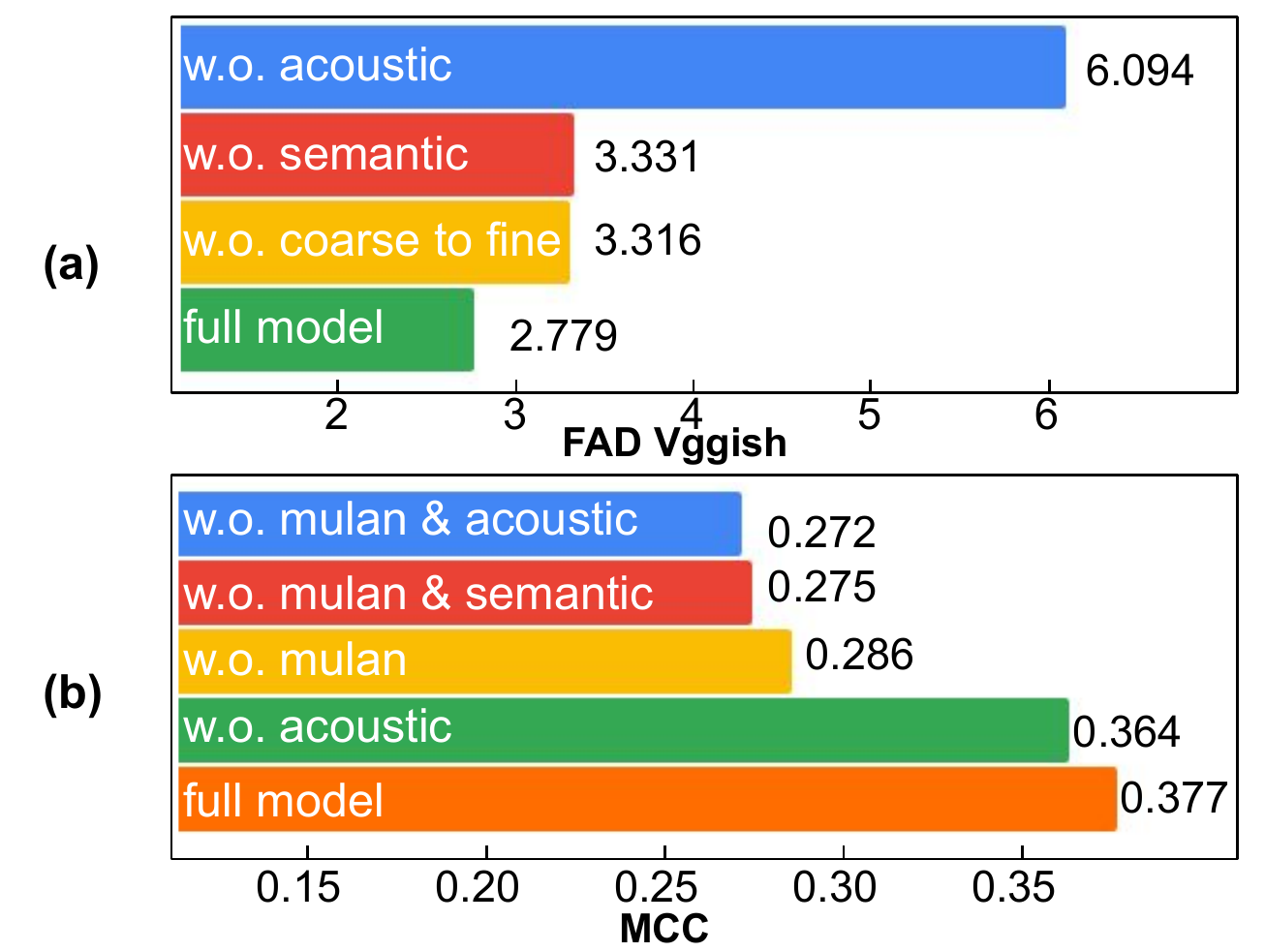}
    \caption{(a) Ablation study on the contribution of each component for MV100K dataset using FAD VGGish score, the lower the better. (b) Ablation study on the contribution of each component of for MusicCaps dataset using MCC score between text and generated audio, the higher the better.}
    \label{fig:ablation}
\end{figure}

Illustrated in Figure~\ref{fig:ablation}, we conducted additional ablation studies to assess the contribution of each stage, (a) on MV100K, measured using FAD VGGish score (↓), and (b) on MusicCaps, measured by MCC score (↑). The results in the figure suggest that video conditioning is crucial for both semantic and acoustic modeling. Furthermore, directly modeling acoustic tokens without semantic tokens proves suboptimal compared to the multi-stage modeling in terms of audio quality and semantic consistency. Directly predicting acoustic tokens from CLIP visual features results in a degradation of the FAD score from 2.779 to 3.331 and a reduction in semantic alignment from 0.377 to 0.275. This novel multi-stage design facilitates the implicit learning of both coarse-grained (style) and fine-grained (rhythm) semantics shared between music and video, enabling the generation of visually relevant soundtracks. Notably, this is achieved despite training solely on 5,000 hours of music videos, where video and audio are not perfectly semantically aligned, unlike in other tasks such as text-to-speech.

\section{Conclusion}
\label{sec:conclusion}
V2Meow is a versatile soundtrack generation model that excels in producing high-fidelity music for diverse video inputs. To overcome the disparity between coarser global video signatures and high-resolution audio representations designed for waveform reconstruction, our approach learns implicit, general video-audio-aligned low-resolution representations. The hierarchical multi-stage autoregressive modeling is crafted to achieve optimal music generation quality. The semantic modeling stage embeds video and audio in a shared semantic space, while the acoustic modeling stage refines the high-resolution audio representation space in a coarse-to-fine fashion. What sets this multi-stage approach apart is its decoupling; only the semantic modeling stage needs to be trained on paired data with semantic video and audio features. The acoustic modeling stage can be trained on audio-only data and optionally fine-tuned on audio data from video to address domain shift. Zero-shot evaluation on dance videos indicates transferability to unseen video input types. Compared to MIDI-based and text-input-only music generation models, V2Meow can generate music more aligned with visual content and human perception. Ablation studies underscore the critical role of video conditioning in both semantic and acoustic modeling stages for generating high-fidelity sounds from video inputs, emphasizing that directly generating acoustic tokens without semantic tokens leads to a degradation in generation quality.

\section{Ethical Statement}
Large generative models learn to imitate patterns and biases inherent in the training set, and in
our case the model might propagate the potential biases built in the video and music corpora used to train our models. Our introduction of text control allows us to debias undesirable stereotypical video-to-music associations. These biases can originate from the video and music corpora used during training, leading to skewed genre distributions and unequal representation of gender, age, and
ethnic groups within each genre. These concerns extend to learned visual-audio associations, which can result in stereotypical links between video content (e.g., people, body movements, dance styles, locations, or objects) and a limited set of musical genres. Additionally, derogatory associations may arise between video choreography and audio output (e.g., minstrelsy, parody, miming).

\section{Acknowledgments}
We are grateful for having the support from Jesse Engel, Ian Simon, Hexiang Hu, Christian Frank, Neil Zeghidour, Andrea Agostinelli, David Ross and authors of MusicLM project for sharing their research insights, tutorials and demos. Many thanks to Austin Tarango, Leo Lipsztein, Fernando Diaz, Renee Shelby, Rida Qadri and Cherish Molezion for reviewing the paper and supplementary materials and share valuable feedbacks regarding responsible AI practice. Thanks Sarvjeet Singh, John Anderson, Hugo Larochelle, Blake Cunningham, Jessica Colnago for supporting publication process. We owe thanks to Muqthar Mohammad, Rama Krishna Devulapalli, Sally Goldman, Yuri Vasilevski, Alena Butryna for supporting our human study. Special thanks to Joe Ng, Zheng Xu, Yu-Siang Wang, Ravi Ganti, Arun Chaganty, Megan Leszczynski, Li Yang for exchanging research ideas and sharing engineering best practice. Thanks Li Li, Jun Wang, Jeff Wang, Bruno Costa, Mukul Gupta for sharing early feedbacks to our demo. Joonseok Lee was partially supported by NRF (2021H1D3A2A03038607, 2022R1C1C1010627, RS-
2023-00222663) and IITP (2022-0-00264).

\bibliography{aaai24}

\section{Appendix}
\subsection{A1.Video Examples}
\begin{figure*}
    \centering
    \includegraphics[width=\linewidth]{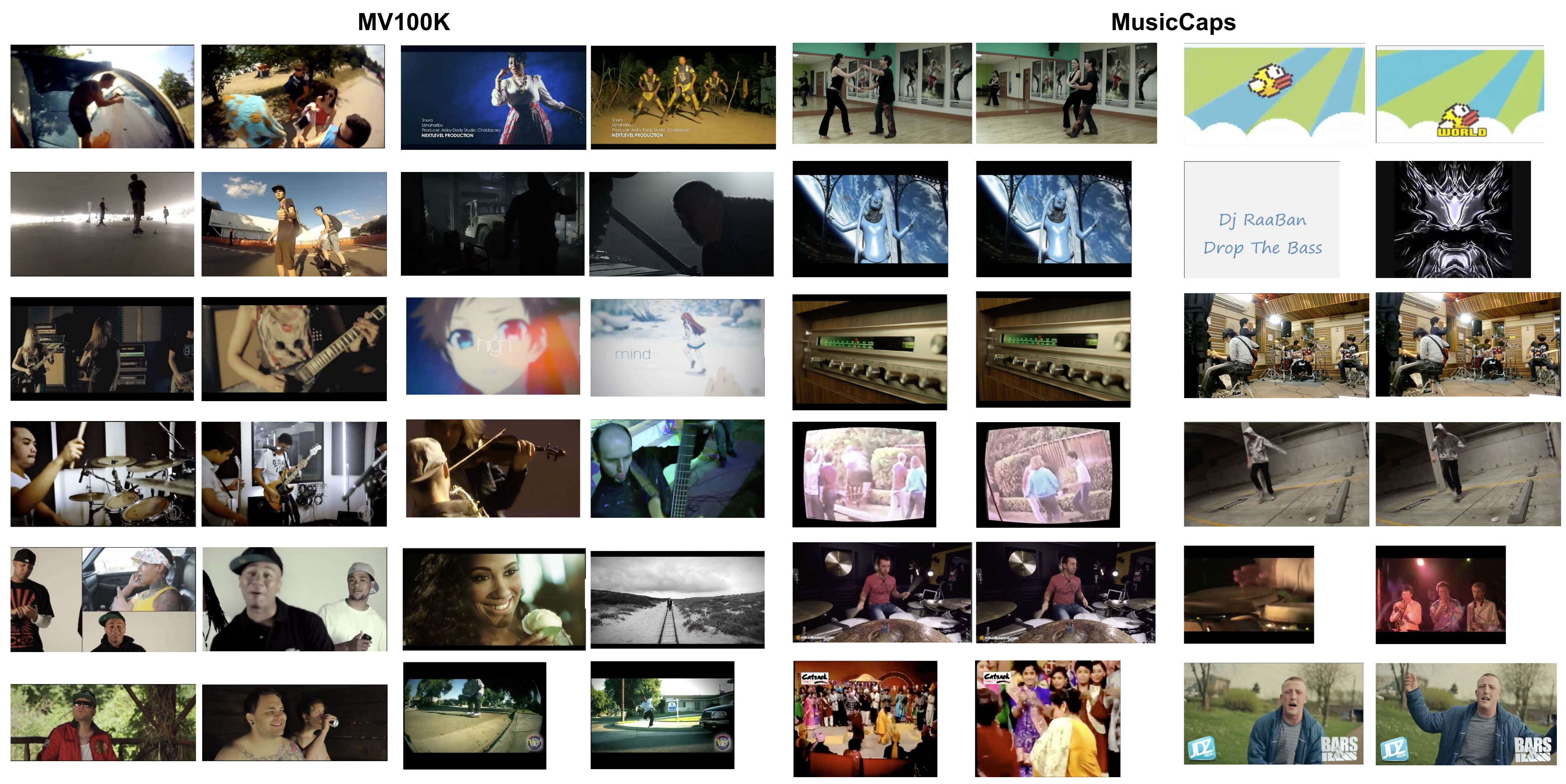}
    \caption{Videos Included in Human Study: (left) Example video inputs included in the MV100K test set. (right) Example video inputs included in the MusicCaps test set. Two frames are shown for each video.}
    \label{fig:video_example}
\end{figure*}

Here we showcase the diverse range of video input types supported by V2Meow models. We demonstrate this with highly rated examples from MV100K and MusicCaps in Figure~\ref{fig:video_example}, which offer videos with both music-related and non-music-related scenes. We classify a scene as music-related if it contains a source of music, such as musical instruments like drums and guitar, live performances, or people singing. Scenes without these elements are not directly related to music, such as video logs of skateboarding, friends hanging out in a camping trip, dance videos, and slide shows of landscape photos. Compared to MV100K, MusicCaps features more dance videos, videos with statics images, or lyric videos, which are sampled from AudioSet\footnote{https://research.google.com/audioset/ontology/music\_1.html}.

Our human study results suggest that V2Meow can handle a wide range of video input types and generate visually relevant soundtracks. For music-related scenes, human raters consider the generated music aligned with the video semantic if it features the sound of the instrument, aligns with the music genre indicated by the combination of these instruments, or features vocals that accompany the movement of facial features. The audio-visual correspondence is abstract rather than physical. The generated music does not need to rigorously follow the exact melody the artist was playing or reproduce precisely what was said or sung to be considered as visually relevant. For non-music-related scenes, if the generated music matches the mood and temporal changes in the scene, it is deemed visually relevant. 

\begin{figure}
    \centering
    \includegraphics[width=\linewidth]{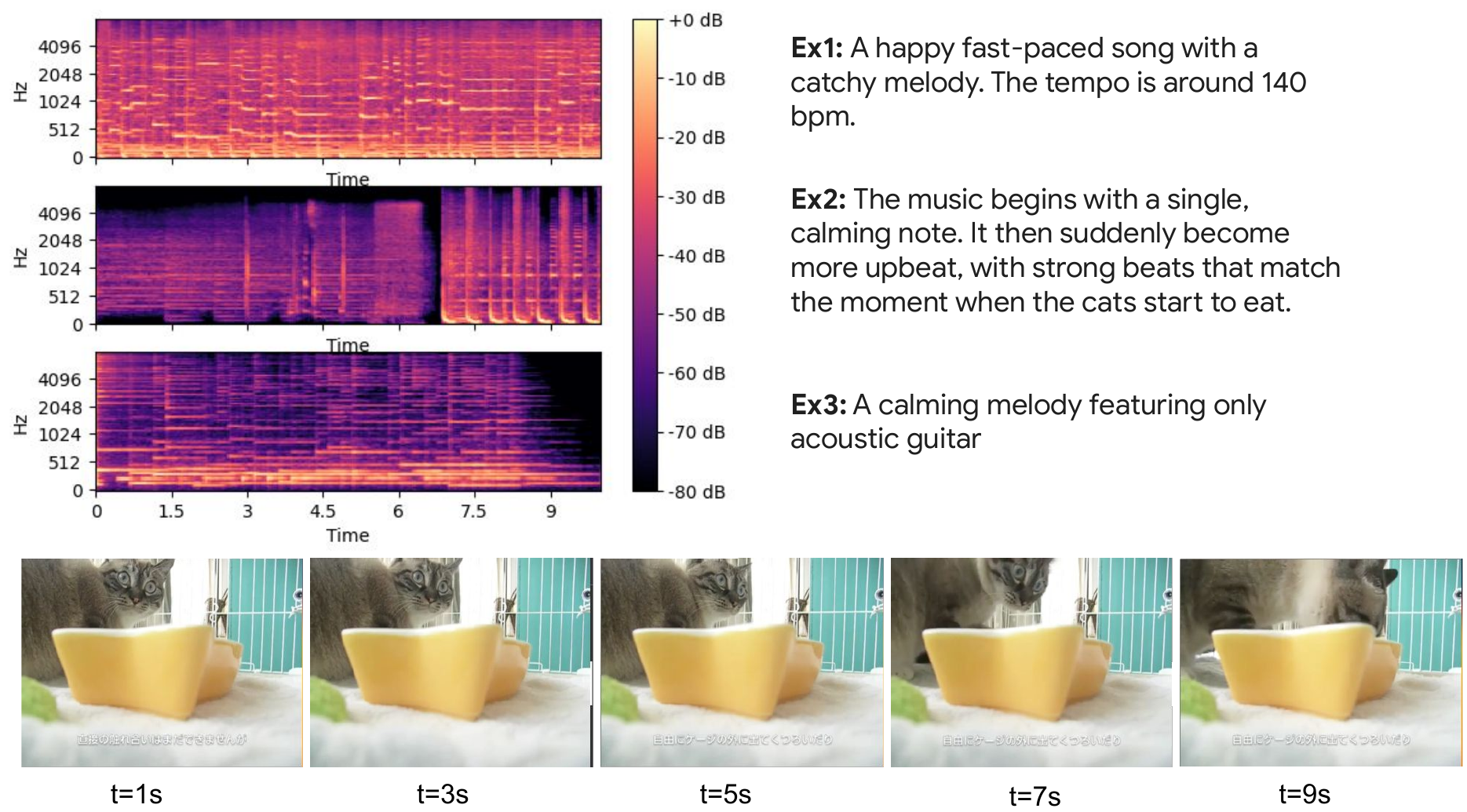}
    \caption{Out-of-Domain Example Analysis: Here we show melspectrograms of 3 music audios generated from the same cat video clip, which features an out-of-domain object, i.e., cat and an audio event, i.e., cat starts to eat at t=7s. }
    \label{fig:cat_example}
\end{figure}

\begin{figure}
    \centering
    \includegraphics[width=\linewidth]{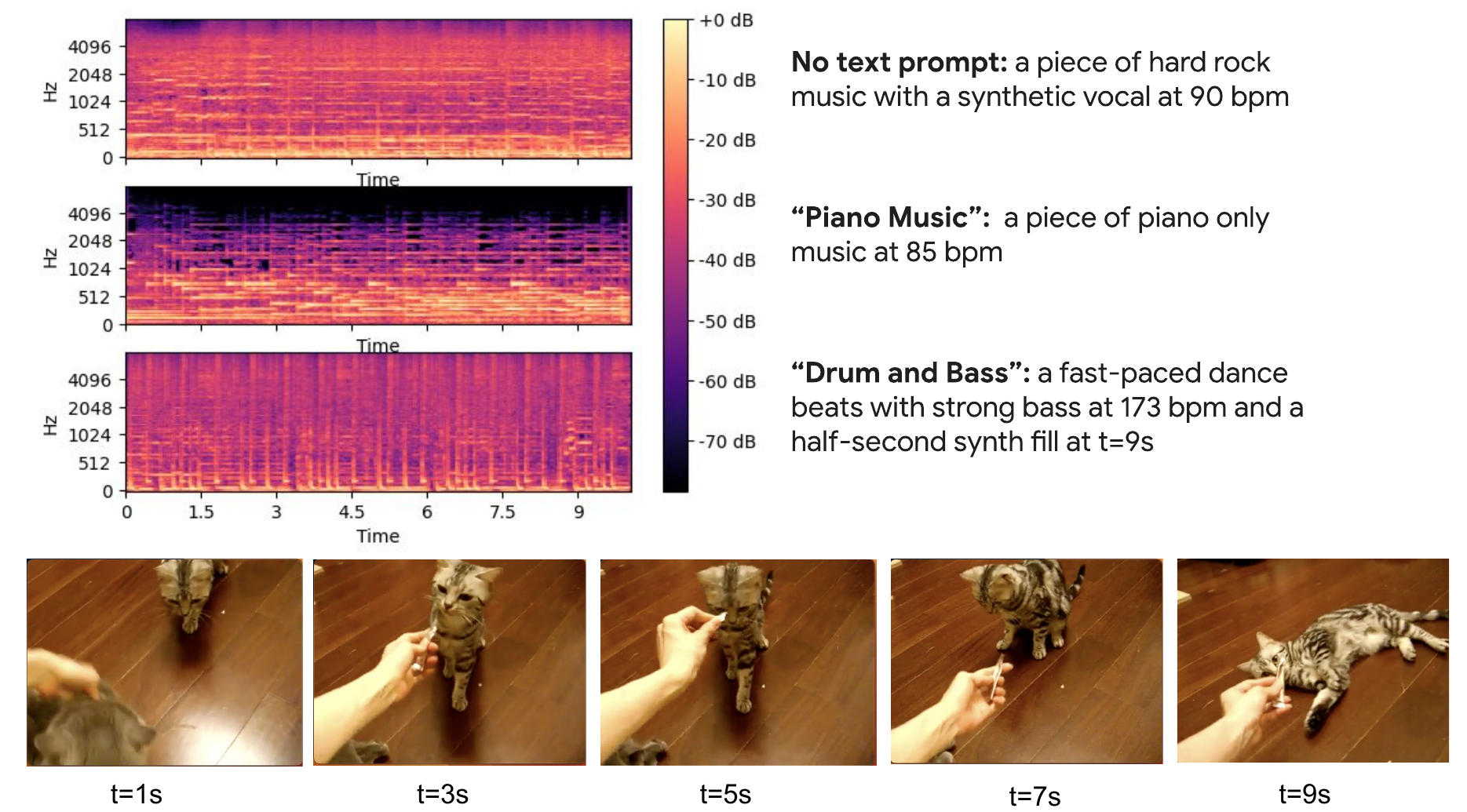}
    \caption{Out-of-Domain Example with Text Control: Here we show melspectrograms of 3 music audios generated from the same cat video clip, with an optional text prompt for control. }
    \label{fig:cat_example_w_text}
\end{figure}

In addition, we evaluate V2Meow's performance on out-of-domain video input types, such as cat videos. Figure~\ref{fig:cat_example} shows three examples of generated music audios from a 10-second cat video clip. The first piece of music is a fast-paced happy song with 140 bpm. The second piece of music features a transition from a calming tune to an upbeat energetic tune, which matches the audio event at t=7s when the cat starts to eat. The third piece is an acoustic guitar solo. The generated music can be controlled via a text prompt as shown in Figure~\ref{fig:cat_example_w_text}. By altering the text prompts, we can add additional control of the generated music on genre, mood, etc., while still being aligned with the video input. For example, with the text prompt ``Drum and Bass'' we can break the undesirable association between the cat video and hard rock music, and generate a soundtrack with dancing beats.

\subsection{A2.Data Details}
\subsubsection{MV100K}
\label{app:yt8m}
For training data we leverage the 100K music videos~\cite{suris2022s} referred here as MV100K. To understand the genre distribution of our training data, we've trained an AudioSet classifier to label the genres of the validation set of size around 22K videos. The multi-label classifier is trained on the music portion of AudioSet data, with a AUC-ROC score of 0.936 on the validation split.

\begin{figure}[!t]
\centering
\includegraphics[width=\linewidth]{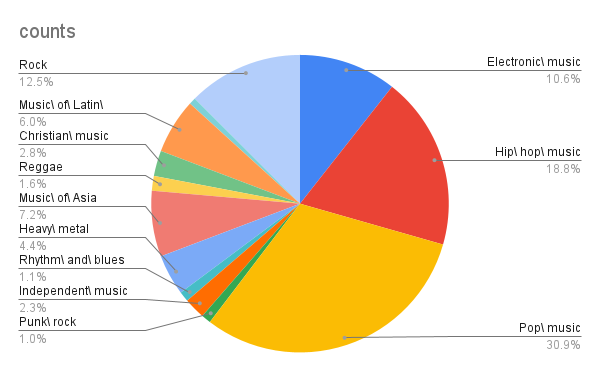}
\caption{Genre distribution of MV100K.}
\label{app:fig:yt8m_genre}
\end{figure}

As shown in Figure~\ref{app:fig:yt8m_genre}, the majority of samples are Pop, Hip-hop and Rock, which is similar to the genre statistics reported by~\cite{suris2022s} on the same MV100K dataset but different in orders. As the train split and validation split share the same distribution, we can conclude that the genre distribution of the training split of MV100K is also skewed and unbalanced. We refer reader to~\cite{suris2022s} for detailed analysis of gender vs genre, race vs genre, age vs genre, visual scene vs genre and instrument vs genre in the training data. For example, per genre analysis has shown that for genre like heavy metal, hard rock, hip hop, it's predominantly male, which may establish a stereotypical relationship between the gender presented in the video input and genre of the generated music. Text-based high-level control can be of great help to handle such bias inherited from the training data.

For numerical evaluation, we selected $4550$ genre-balanced 10-seconds video clips from the MV100K validation split. The genre distribution can be found in Table~\ref{app:table:dataset}. For human study we further sub-sampled $89$ distinct examples from $9$ different genres. Note that ``Rock'' genre in AudioSet is a parent genre for Punk rock and Heavy metal. In this study, Rock genre represents all music clips that are not Punk rock or Heavy metal. 

\begin{table*}[]
\centering
\small
{%
\begin{tabular}{l|c|c|c}
\toprule
Genre Mid & Genre Name & No. of Examples & In Human Study\\
\midrule
/m/02lkt            & Electronic\textbackslash music                                   & 350                                          & YES \\
/m/0glt670          & Hip\textbackslash hop\textbackslash music                        & 350                                          & YES  \\
/m/064t9            & Pop\textbackslash music                                          & 350                                          & YES  \\
/m/05r6t            & Punk\textbackslash rock                                          & 350                                          & YES  \\
/m/05rwpb           & Independent\textbackslash music                                  & 350                                          & NO   \\
/m/06j6l            & Rhythm\textbackslash and\textbackslash blues                     & 350                                          & NO  \\
/m/03lty            & Heavy\textbackslash metal                                        & 350                                          & YES \\
/m/028sqc           & Music\textbackslash of\textbackslash Asia                        & 350                                          & YES  \\
/m/06cqb            & Reggae                                                           & 350                                          & YES  \\
/m/02mscn           & Christian\textbackslash music                                    & 350                                          & NO  \\
/m/0g293            & Music\textbackslash of\textbackslash Latin\textbackslash America & 350                                          & NO   \\
/m/0164x2           & Music\textbackslash of\textbackslash Africa                      & 350                                          & YES  \\
/m/06by7            & Rock                                                             & 350                                          & YES  \\ \bottomrule
\end{tabular}%
}
\caption{Genre distributions of MV100K used in numerical study.}
\label{app:table:dataset}
\end{table*}

\subsubsection{MusicCaps}
\label{app:mc}
MusicCaps is a music caption dataset derived from AudioSet, where it contains $5,521$ audio clips with detailed description of the music audio. Here we augment the dataset with 10-s video clip. The genre distribution can be found in~\cite{Agostinelli2023MusicLMGM}. For human study, we sampled $100$ genre-balanced examples from the MusicCaps.

\subsection{A3.Human Study Details}
\label{app:human}
\begin{figure*}[!t]
    \centering
    \includegraphics[width=\textwidth]{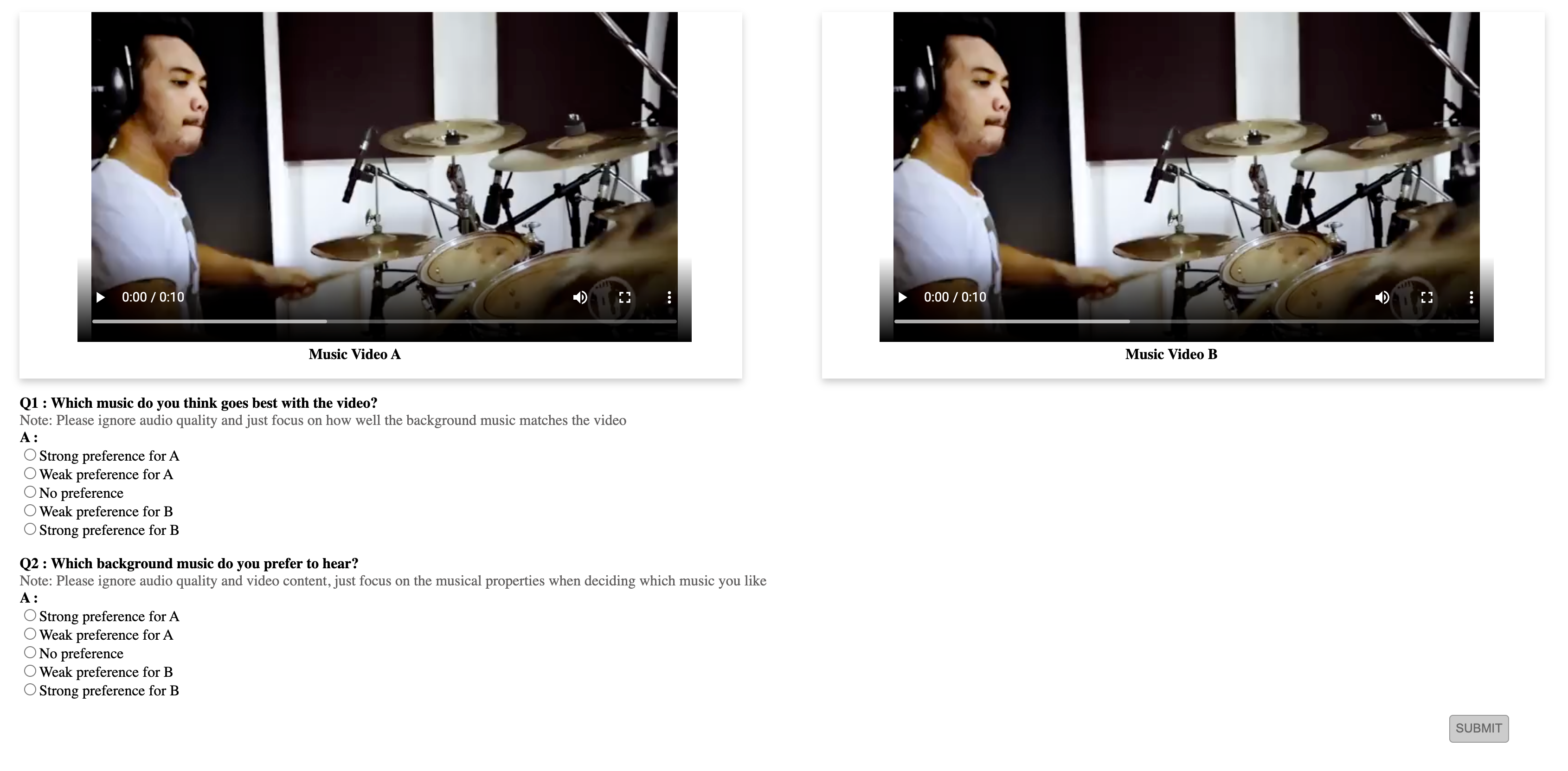}
    \caption{The UI for Human Study.}
    \label{fig:human_study_ui}
\end{figure*}

As shown in Figure~\ref{fig:human_study_ui}, around $200$ participants in the listening test were presented with two 10-second clips that have different background music, and asked to compare two background music in terms of 1) visual relevance 2) music preference on a 5-point Likert scale. Each pair of radio is rated by $3$ raters. The order of the video clips shown to the listeners is randomized to prevent biased ratings. 

\begin{figure*}[!t]
\centering
\includegraphics[width=\linewidth]{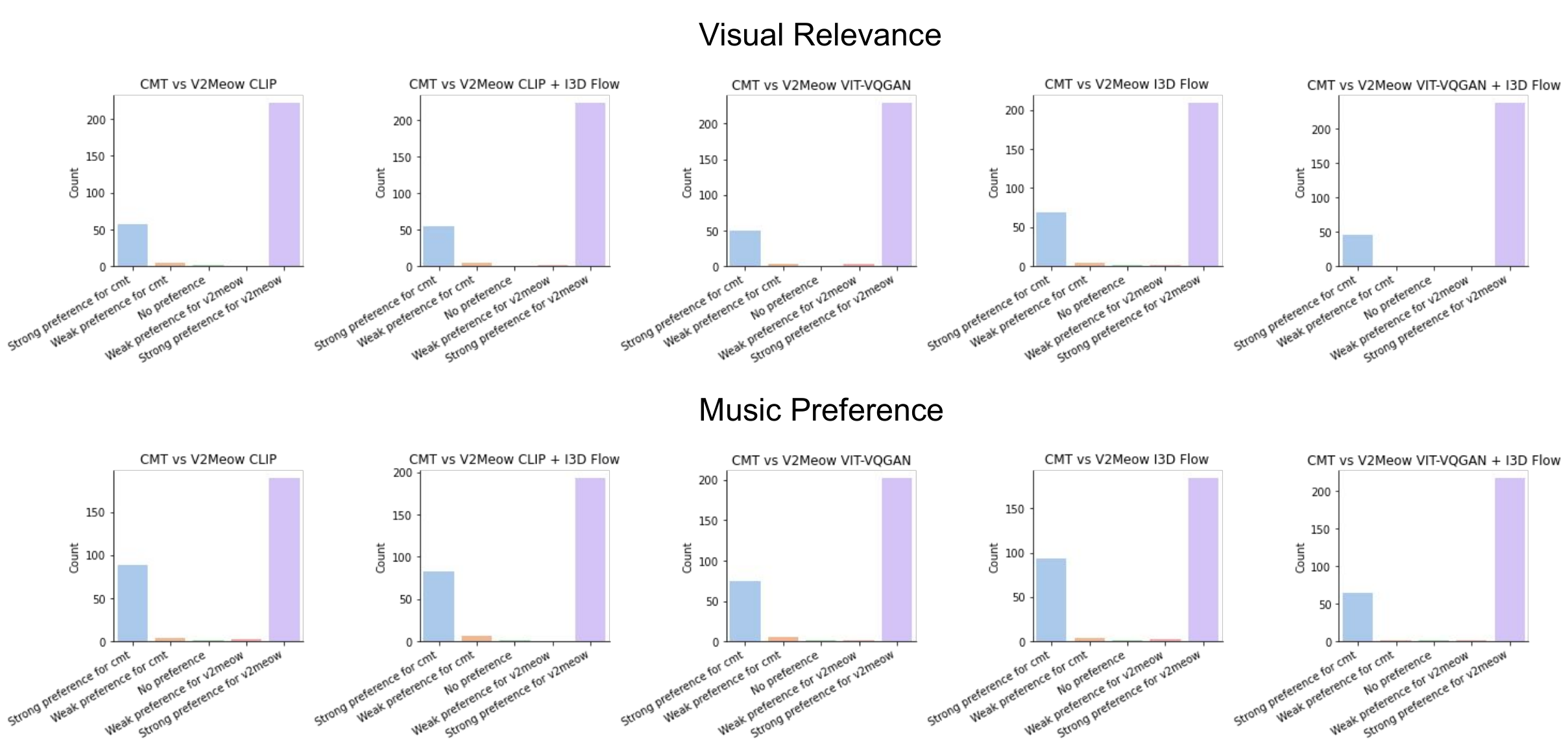}
\caption{Pairwise comparisons of visual relevance and music preference from the human listener study on MV100K. Each pair is compared on a 5-point Likert scale.}
\label{app:fig:mv100_human_study}
\end{figure*}

For the MV100K human study results shown in Table~\ref{tab:quantitative results}, we compare each of the $5$ video-to-music variants with baseline model CMT on $89$ distinct video examples from MV100K, and collected $1424$ ratings. The detailed pairwise comparison between V2Meow models and CMT can be found in Figure~\ref{app:fig:mv100_human_study}. 

\begin{figure*}[!t]
\centering
\vspace{0.5cm}
\includegraphics[width=\linewidth]{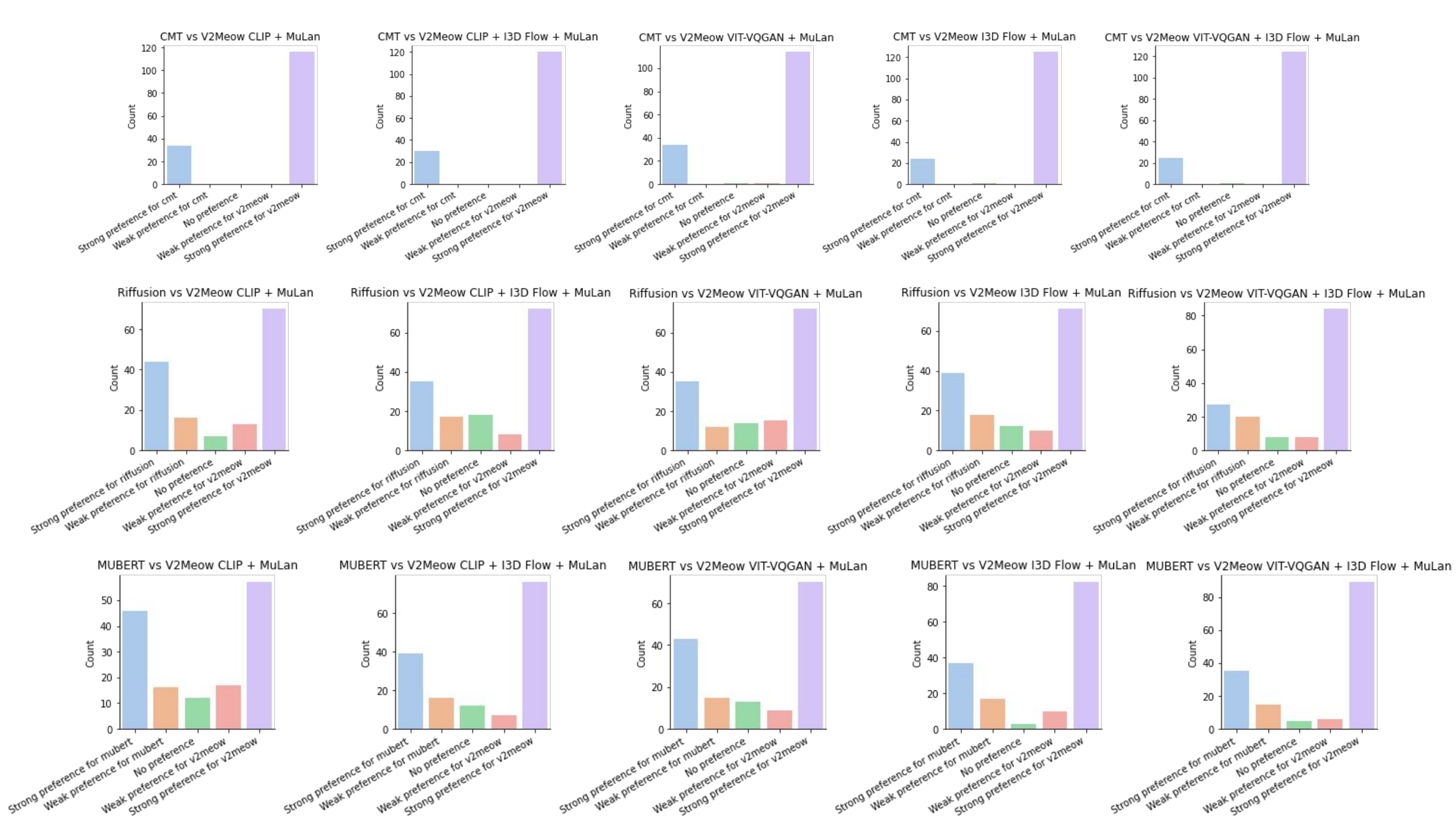}
\caption{Pairwise comparisons of visual relevance from the human listener study on MusicCaps. Each pair is compared on a 5-point Likert scale.}
\label{app:fig:musiccaps_visual_relevance}
\end{figure*}

\begin{figure*}[!t]
\centering
\includegraphics[width=\linewidth]{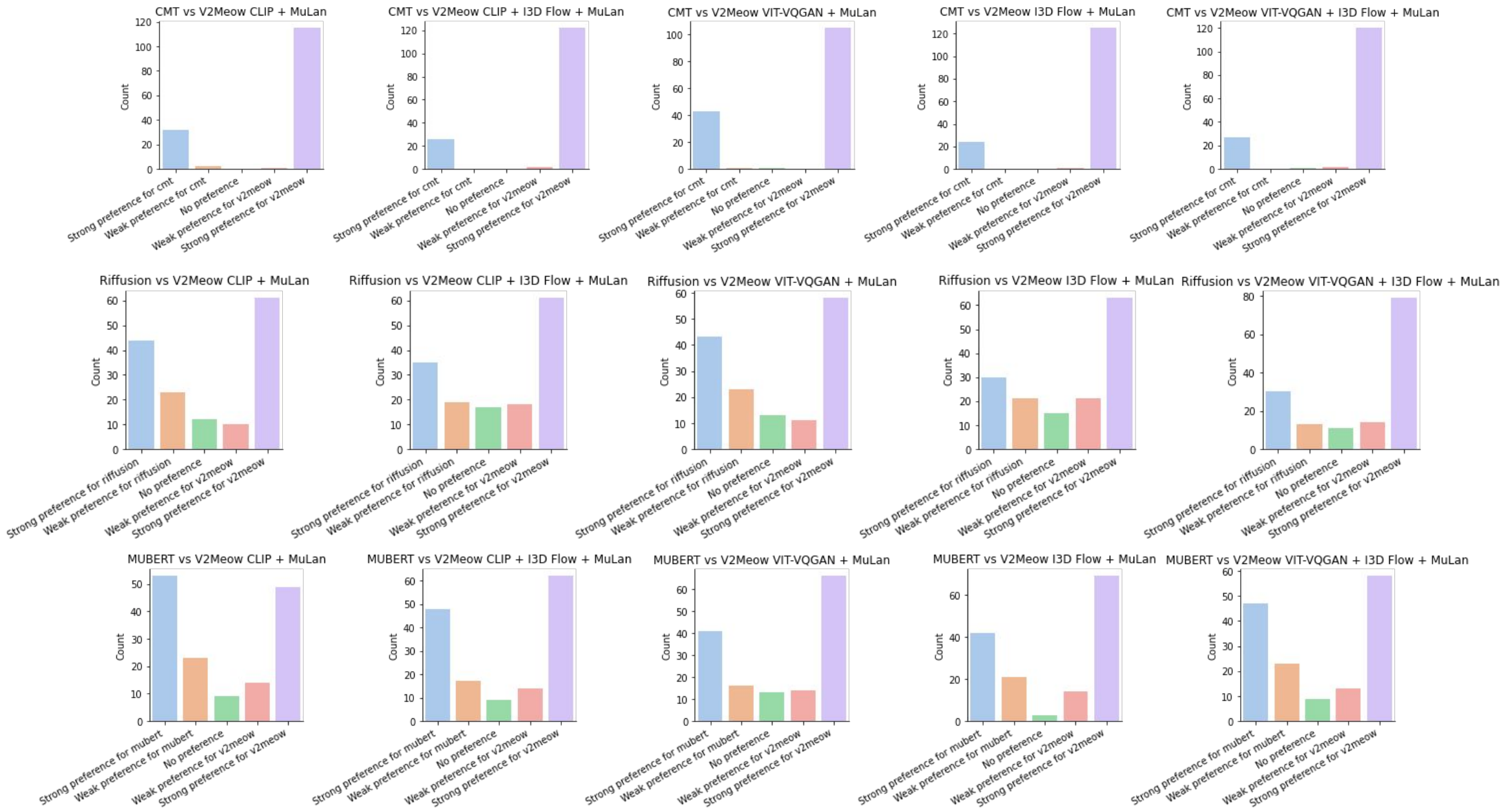}
\caption{Pairwise comparisons of music preference from the human listener study on MusicCaps. Each pair is compared on a 5-point Likert scale.}
\label{app:fig:musiccaps_music_quality}
\end{figure*}

For the MusicCaps human study results shown in Table~\ref{tab:quantitative results}, we compare each of the $5$ video-text-to-music variants with baseline model CMT, mubert and riffusion on $76$ distinct video examples from the MusicCaps dataset. For each pair of models, we sample $50$ examples randomly. As each example is rated by $3$ raters, in the end we've collected $2392$ ratings. The detailed pairwise comparison results for visual relevance is shown in Figure~\ref{app:fig:musiccaps_visual_relevance} and results for music quality is shown in Figure~\ref{app:fig:musiccaps_music_quality}.

\begin{figure}[!t]
\centering
\includegraphics[width=\linewidth]{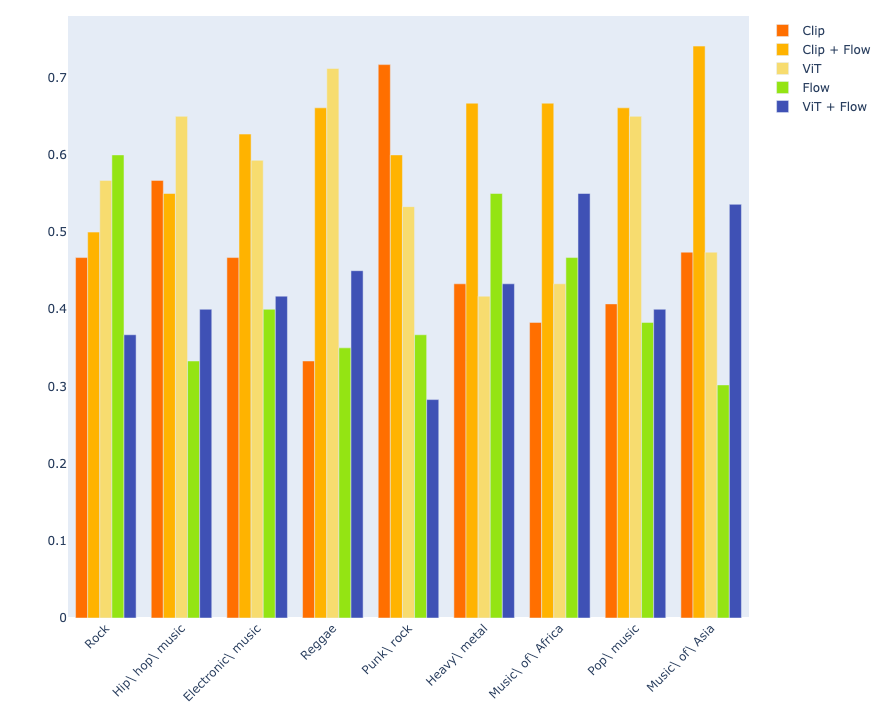}
\caption{Model winning rate per genre for visual relevance task.}
\label{app:fig:genre_task1}
\end{figure}

\subsection{A4.Ablation Study and Genre Analysis}
\label{app:genre_human}

\begin{table}[]
\centering
\small
{%
\begin{tabular}{l|cc}
\toprule
Base Model\textbackslash{}Component & Visual Relevance &  Music Preference \\
\midrule
CLIP  & 47.2\%                                 & 49.0\% \\
CLIP + I3D Flow   & 63.0\%                                 & 61.8\%                             \\
VIT-VQGAN   & 55.9\%                                 & 53.6\%                             \\
I3D Flow & 41.7\%                                 & 40.7\%                             \\
VIT-VQGAN + I3D Flow & 42.6\%                                 & 45.3\%                            \\ \bottomrule
\end{tabular}%
}
\caption{Ablation study on human perceptual evaluation. The value indicates the percentage of people who select the method. The statistics is compute from around $1409$ ratings.}
\label{tab:human_eval_ablation}
\end{table}

We've also run ablation study on MV100K to compare different visual features and their impacts on visual relevance and music quality. For each genre and each pair of video-to-music model variants (a total of $10$ pairwise combination were made from the $5$ available model variants), we randomly sample $5$ examples resulting in $50$ pairs of generated soundtrack per genre and around $150$ ratings per genre for pairwise comparison. The total number of pairwise ratings is $1409$. The winning rate per genre is calculated as the ratio of the number of Strong or Weak Preference out of $150$ ratings. The average winning rate is the winning rate per genre averaged over $9$ genres. 

As shown in Table~\ref{tab:human_eval_ablation}, Clip+flow and ViT model have been the top 2 preferred model if measured by the average winning rate for both the visual relevance and music quality task. As shown in the per genre analysis in Figure~\ref{app:fig:genre_task1}, Clip+Flow model is also the best performing model for most genres especially Music of Asia, a genre features predominantly dance videos like KPop videos. While ViT model performs best for Reggae. For most genres, the combination of Clip and Flow results in better performance, as Clip tends to capture semantic context and Flow tends to capture temporal context. However, in the Punk Rock genre Clip model performs better than the Clip+Flow model, which can be explained by the stronger connection between the semantic context of the video and the Punk Rock genre and weaker connection between the temporal context.

\end{document}